\documentclass[preprint2]{aastex}

\newcommand{\ee}{$e^\pm$}
\newcommand{\g}{$\gamma$}
\newcommand{\wfe}{W_{{\rm K}\alpha}}
\newcommand{\nh}{N_{\rm H}}
\newcommand{\sax}{{\it BeppoSAX}}
\newcommand{\xte}{{\it RXTE}}
\newcommand{\exosat}{{\it EXOSAT}}
\newcommand{\rosat}{{\it ROSAT}}
\newcommand{\asca}{{\it ASCA}}
\newcommand{\ginga}{{\it Ginga}}
\newcommand{\ec}{E_{\rm c}}
\newcommand{\eb}{E_{\rm b}}
\newcommand{\efe}{E_{{\rm K}\alpha}}
\newcommand{\sfe}{\sigma_{{\rm K}\alpha}}
\newcommand{\ife}{I_{{\rm K}\alpha}}
\newcommand{\cnu}{\chi^2_\nu}

\begin{document}

\title{The broad-band spectrum of 3C 120 observed by BeppoSAX}

\author{Andrzej A. Zdziarski}
\affil{N. Copernicus Astronomical Center, Bartycka 18, 00-716 Warsaw, Poland}
\email{aaz@camk.edu.pl}
\and
\author{Paola Grandi}
\affil{Istituto di Astrofisica Spaziale, CNR, Via Fosso del Cavaliere 100,
I-00133 Roma, Italy}
\email{paola@ias.rm.cnr.it}

\begin{abstract} We study a broad-band X-ray observation by \sax\/ of the
broad-line radio galaxy 3C 120. The primary X-ray continuum is well described
by a power law with the photon index of $\Gamma\sim 1.85\pm 0.05$ and a
high-energy break or a cutoff. The obtained e-folding energy of $\sim 100$--300
keV corresponds to $kT\sim 50$ keV in thermal-Compton models. A soft X-ray
excess is found at a statistical significance of $\ga 98\%$. Our physical
best-fit model of the excess is optically-thin emission from an extended plasma
(with a luminosity of $\sim 2\%$ of the total X-ray luminosity), which
interpretation is supported by an independent finding of an extended soft X-ray
halo in 3C 120. We find a moderate Compton reflection component together with a
moderately broad Fe K$\alpha$ line with an equivalent width ($\sim 10^2$ eV)
fully consistent with the strength of reflection, indicating their origin in an
optically-thick accretion disk extending relatively close to the central black
hole. We observe strong spectral variability during our 2-day observation with
$\Gamma$ varying from $\sim 1.7$ to $\sim 2$ and correlated with the soft X-ray
flux. The variability is very well modeled by thermal Comptonization in a hot
plasma irradiated by a variable UV flux. Both a hot patchy corona outflowing
with a mildly relativistic velocity away from a cold disk and a hot inner
accretion flow overlapping with the cold disk are viable geometries. The
presence of an outflow in the first case and an overlap between the hot and
cold phases in the second case are required by energy balance and the observed
strength of Compton reflection. \end{abstract}


\section{INTRODUCTION}
\label{s:intro}

3C 120 is a broad-line radio galaxy (BLRG) at $z=0.033$ exhibiting
characteristics intermediate between those of FR I galaxies and BL Lacs (Urry
\& Padovani 1995). It is also classified as a Seyfert 1 (Burbidge 1967),
although its optical morphology is complex, see Grandi et al.\ (1997).  It is
variable in all energy bands, and shows superluminal motion in the
VLBI core (e.g., Walker 1997).

In 1994, 3C 120 has been observed by \asca, and since then the strength and the
width of its Fe K$\alpha$ has been a subject of a considerable controversy.
Reynolds (1997) found the line to be extremely strong, with an equivalent width
of $\wfe\approx 1$ keV, as well as extremely broad, with a Gaussian width of
$\sfe\approx 2$ keV. The physical origin of such a line is unclear. A slightly
weaker line, with $\wfe\approx 0.5$ keV and $\sfe\sim 1$ keV, was found in the
same data by Grandi et al.\ (1997). Both results were obtained assuming the
underlying continuum was a single power law in the entire energy range observed
by \asca, and Grandi et al.\ (1997) mentioned the possibility that the unusual
line parameters  were caused by that assumption not being valid. Indeed,
Wo\'zniak et al.\ (1998, hereafter W98) found that if the \asca\/ data (with an
updated calibration of 1997) were fitted in the 4--10 keV range assuming the
presence of Compton reflection with a reflector solid angle of $\Omega$ between
0 and $2\pi$, the parameters of the line became much more physically plausible,
with the best-fit values of $\wfe\approx 60$--110 eV, the peak energy, $\efe
\approx 6.4$ keV, and $\sfe\simeq 0.32$ keV and 0.20 keV assuming $\Omega=0$
and $2\pi$, respectively. The line in the \asca\/ data was again claimed to be
very strong and broad by Sambruna, Eracleous \& Mushotzky (1999), who obtained
$\wfe=510_{- 90}^{+70}$ eV and $\sfe=0.89^{+0.17}_{- 0.15}$ keV, virtually
identical with the values of Grandi et al.\ (1997).

This issue has recently been resolved thanks to results of \xte\/ and \sax.
Eracleous, Sambruna \& Mushotzky (2000, hereafter E00) have found the line in
the \xte\/ data ($\geq 4$ keV) to be weak with the parameters consistent with
those of W98, namely $\wfe=90^{+30}_{- 20}$ eV and $\sfe<1.2$ keV. Similarly,
Grandi (2000, hereafter G00) have found $\wfe=60^{+80}_{- 40}$ eV and
$\sfe=0.27_{-0.27}^{+0.40}$ keV in a preliminary analysis of the \sax\/ data.

Since \ginga\/ did not observe 3C 120, the actual strength of Compton
reflection had been unknown until the results of E00 and G00, who found
$\Omega/2\pi=0.4^{+0.4}_{- 0.1}$ and  $0.7^{+0.4}_{-0.4}$, respectively. These
values of the solid angle are compatible (George \& Fabian 1991; \.Zycki \&
Czerny 1994) with the presence of a modest Fe K$\alpha$ line with the
parameters of W98, E00 and G00.

Here, we present a comprehensive analysis of the observation of 3C 120 by \sax,
which combines the broad-band hard X-ray coverage of \xte\/ with a high energy
resolution and an energy range in soft X-rays comparable to that of \asca. We
address the issues of the parameters of the line and the continuum, their
variability, radiative processes and geometry of the source, and the
relationship of 3C 120 to other BLRGs as well as Seyferts.

\section{DATA REDUCTION}
\label{s:reduction}

The observation log is given in Table \ref{t:log}. Data from each instrument
have been reduced following the standard procedures (Fiore, Guainazzi \& Grandi
1999, hereafter F99). Spectra of LECS and MECS (2 units) detectors were
extracted using a radius of $6'$ and $4'$, respectively. The background spectra
were taken from the blank sky fields in the position of the source with the
same extraction regions.

\begin{deluxetable}{lccccccc}
\tabletypesize{\small}
\tablecolumns{8}
\tablecaption{The observation log}
\tablehead{\colhead{Start time} &\colhead{End time}
&\multicolumn{3}{c}{Exposure [s]}
&\multicolumn{3}{c}{Count rate [s$^{-1}$]\tablenotemark{a}}\\
\colhead{} & \colhead{} &\colhead{LECS} &\colhead{MECS} &\colhead{PDS}
&\colhead{LECS} &\colhead{MECS}
&\colhead{PDS} }
\startdata
1997-09-20 08:14:15 &1997-09-22 16:22:34 &34029 &77758 &35755 &$0.368\pm 0.003$
& $0.638\pm 0.003$ &$0.72\pm 0.03$ \\
\enddata
\tablenotetext{a}{The energy ranges are 0.3--4 keV, 1.3--10 keV, 13--110 keV
for LECS, MECS, PDS, respectively, the count rates are background subtracted,
and their uncertainties are 1-$\sigma$.
}
\label{t:log}
\end{deluxetable}

We bin the LECS and MECS spectral files proportionally to the instrumental
resolution (F99). After rebinning, there is always more than 20 counts per
channel, as required for the validity of $\chi^2$ fitting. The LECS data are
analyzed in the 0.3--4 keV energy range (channels 24--80). The lower limit is
above the standard value of 0.15 keV (F99) because the source was strongly
dominated by the background below 0.3 keV. The energy range of 1.3--10 keV
(channels 34--114) is used for the MECS data. We allow a free normalization of
the LECS spectrum with respect to the MECS one (F99).

We use the variable rise-time PDS spectrum. This choice is in agreement with
F99, who recommend this method for weak sources with count rates $\la 0.5$--1
s$^{-1}$ and the statistical uncertainty of the photon index, $\Gamma$, of
$\Delta \Gamma\ga 0.1$, and with spectra similar to that of the Crab. Indeed,
$\Gamma\simeq 2.1 \pm 0.1$ in the PDS data, very similar to the Crab spectrum,
and the 13--110 keV count rate is 0.72 s$^{-1}$. After using the logarithmic
rebinning in 18 channels (F99), photons are detected up to channel 15, and
hereafter we use channels 2--15. When the variable and fixed rise-time spectra
are fitted by the same model, the normalization of the former is about 0.95 of
the latter. The same value is obtained by directly dividing the two spectra in
the 15--100 keV energy range. If the lowest channels (13--15 keV) are also
included, the ratio becomes slightly lower (0.93). From this and
recommendations of F99, we derive the allowed relative normalization of the PDS
spectrum with respect to the MECS one of 0.77--0.85.

\section{SPECTRAL ANALYSIS}

\subsection{Spectral Models}
\label{s:models}

We use {\sc xspec} (Arnaud 1996) for spectral fitting. The parameter
uncertainties below correspond to 90\%  confidence for a single parameter,
i.e., $\Delta \chi^2=+2.7$, the plotted error bars are 1-$\sigma$, and the
displayed spectral data are rebinned for clarity.  The significance level of a
hypothesis is obtained, unless stated otherwise, from the $F$-test. For
calculations of luminosity, we assume $H_0=75$ km s$^{-1}$ Mpc$^{-1}$, which
gives the distance of 134  Mpc.

Intrinsic model spectra are absorbed by the interstellar medium in the Galaxy
with a column density of $N_{\rm H}^{\rm G}=1.23\times 10^{21}$ cm$^{-2}$
(Elvis, Lockman \& Wilkes 1989). We also allow for intrinsic absorption at the
redshift of the source with a column, $\nh$. The absorption is in neutral
matter with the abundances of Anders \& Ebihara (1982) and the opacities of
Morrison \& McCammon (1983).

We allow for Compton reflection of the primary continuum from an
optically-thick medium. We use Green's functions for angle-dependent reflection
of isotropic incident radiation of Magdziarz \& Zdziarski (1995). The
inclination of 3C 120 is most likely close to face-on (Zenzus 1989), and we
thus use $\cos i=0.95$, which is the maximum value for which the Green's
functions are given. The relative amount of reflection is measured by the solid
angle, $\Omega$, subtended by the reflector. The opacities of reflector are the
same as for the absorber. We also test whether the reflecting medium can be
ionized, in which case we use the one-zone ionization  model of Done et al.\
(1992) modified as in Gondek et al.\ (1996). We assume then a reflector
temperature of $10^5$ K (Krolik \& Kallman 1984). The ionization parameter is
defined as $\xi\equiv L_{\rm ion}/n d^2$, where $L_{\rm ion}$ is the luminosity
in an incident power-law continuum in the 5 eV--20 keV range, $n$ is the
density, and $d$ the distance between the source of radiation and the
reflector.

Reflection of the continuum is accompanied by emission of fluorescent lines,
the most prominent of those is the Fe K$\alpha$ line. In models with static
reflection (as above), we model the line as a Gaussian, in which we constrain
the line width to $\sfe\leq 1$ keV (so as to avoid this component to play a
role of an additional continuum).  The line equivalent width, $\wfe$, is
proportional to $\Omega$ at a given $\Gamma$ (George \& Fabian 1991; \.Zycki \&
Czerny 1994). In the case of an optically-thick accretion disk 
with a low ionization, cosmic abundances, an inclination of
$i\sim 0$, and the photon index of $\Gamma\simeq 1.8$--1.9 
(which we find {\it a
posteriori\/} to be the case for our data, see \S \ref{s:results}),
calculations of George \& Fabian (1991) give approximately
\begin{equation}
\wfe={\Omega\over 2\pi} 150\, {\rm eV}, \label{eq:line}\end{equation}
which condition we impose in our models. 
On the other hand, a remote
molecular torus may be also present. It is likely that $\nh$ of the torus
is substantially less than $10^{24}$ cm$^{-2}$, as in most of 
Seyfert 2s (e.g.\
Risaliti, Maiolino \& Salvati 1999) and
both narrow and broad line radio galaxies (Sambruna et al.\ 1999), 
as well as it is the case on average for
Seyfert 1s (Lubi\'nski \& Zdziarski 2001). In this case, reflection from the
torus will be negligible (e.g., W98) but a a narrow Fe K$\alpha$ line at 6.4
keV will still be produced (e.g., Makishima 1986). Thus, we allow the presence
of a line with $\efe=6.4$ keV and $\sfe=0$ keV and a free normalization in
addition to that tied to the reflection. 

When the reflection comes from a fast rotating disk in a strong gravitational
potential, Doppler and gravitational shifts become important. We approximate
these effects by convolving {\it both\/} the reflected component and the Fe
K$\alpha$ line (with $\efe=6.4$ keV and $\sfe=0$ keV in the rest frame) with
the Schwarzschild disk profile of Fabian et al.\ (1989). In calculating the
relativistic distortion, we consider a range of radii from $r_{\rm in}\geq 6$
to $r_{\rm out} = 1000$, where $r$ is in units of the gravitational radius,
$R_{\rm g} \equiv GM/c^2$. We assume the emissivity of the reflection and the
line to vary along the disk radius in the same way as the accretion disk
emissivity, as expected when X-ray emitting regions above the disk derive their
power from accretion. To properly describe the line in this detailed model, we
also include the component of the Fe K$\alpha$ line singly-scattered in the
disk at 6.2 keV containing 12\% of the photons in the main component, as well
as the Fe K$\beta$ and Ni K$\alpha$ components, the sum of which we model as a
single disk-line at 7.2 keV containing 18\% of the photons in the main
component (see W98). In those models, $\wfe$ is defined for the Fe K$\alpha$
component only (including the downscattered part), and it is tied to $\Omega$
via equation (\ref{eq:line}).

\subsection{Results}
\label{s:results}

The spectrum is rather poorly fitted by a power-law continuum ($\Gamma=1.82$)
absorbed by $N_{\rm H}^{\rm G}$, at $\cnu=256.4/148$. We have found that the
bad fit is clearly caused by the  presence of soft-excess, Fe K$\alpha$,
Compton-reflection, and high-energy cutoff spectral features.

We then fit an intrinsic continuum of an e-folded power law accompanied by
static Compton reflection and a Gaussian line. This model provides a fair fit
to the data, with $\Gamma=1.96^{+0.05}_{-0.05}$, $\Omega/2\pi=
0.96^{+0.38}_{-0.32}$, $\ec= 250^{+770}_{-120}$ keV, $\nh=
7.7^{+1.9}_{-1.7}\times 10^{20}$ cm$^{-2}$, at $\cnu=159.4/142$. However, there
is a strong systematic residual feature at $\sim 1$ keV  as well as the line
width is at its allowed upper limit, $\sfe=1$ keV, which is not physically
consistent with the assumed static reflection. If we let the line flux to be
independent of reflection, $\sfe<1$ keV, but then $\wfe$ is only $\sim 0.4$ of
the value expected from the fitted reflection (see discussion below on the
absence of resonant absorption), as well as there is no improvement of
$\chi^2$. Any possible contribution to the reprocessing features by the torus
would only strengthen this discrepancy as a torus can produce a line without
Compton reflection (see \S \ref{s:models} above).

We find that allowing for an additional complexity of the spectrum in the soft
X-ray range significantly improves the fit as well as makes the model more
physically consistent. We first allow for a soft power-law component. This
improves $\chi^2$ to 152.9/140 (corresponding to 95\% significance). The
additional power law is quite soft, $\Gamma=4.6^{+0.9}_{-1.6}$, and a rather
large $\nh= 3.2^{+1.3}_{-2.1}\times 10^{21}$ cm$^{-2}$ is required. Then, we
consider an incident e-folded broken power law. We indeed find a break energy
at $\eb=1.0$ keV. However, somewhat surprisingly, we find that the best-fit
model has a power law index, $\Gamma_1=1.00$, below $\eb$ {\it harder\/} than
that, $\Gamma_2=1.88$, above $\eb$. The best-fit $\nh$ is 0 now, which explains
the presence of the soft excess in the previous single power-law models with
$\nh\sim 10^{21}$ cm$^{- 2}$. The present model yields $\ec=170$ keV,
$\Omega/2\pi= 0.61$, $\sfe=0.40$ keV, null flux in the narrow line not
associated with reflection, and $\cnu= 150.5/140$ (the $\chi^2$ reduction is
significant at 98\%). However, as we discuss below, such a broken power law
does not appear to correspond to a physical model.

\begin{deluxetable}{lccccccccc}
\tabletypesize{\small}
\tablewidth{0pc}
\tablecolumns{10}
\tablecaption{Parameters of the model with an {\rm e}-folded power law,
reflection tied to the F{\rm e} K$\alpha$ line and soft thermal emission}
\tablehead{ \colhead{$\nh$\tablenotemark{a}} &\colhead{$\Gamma$}
&\colhead{$\Omega/2\pi$} &\colhead{$\ec\,$[keV]}
&\colhead{$\efe\,$[keV]} &\colhead{$\sfe\,$[keV]} &\colhead{$\wfe\,$[eV]}
&\colhead{$kT_{\rm s}\,$[keV]} &\colhead{$I_{\rm
s}$\tablenotemark{b}}  &\colhead{$\cnu$} }
\startdata
$6.8^{+2.5}_{-2.3}$ & $1.86^{+0.07}_{-0.06}$ & $0.56_{-0.20}^{+0.44}$ &
$150_{-30}^{+230}$ & $6.25_{-0.40}^{+0.24}$ & $0.40_{-0.18}^{+0.60}$ &
$84_{-30}^{+66}$ & $1.06_{-0.27}^{+0.44}$ & $6.1_{-3.3}^{+5.3}$ &
150.5/141\\
\enddata
\tablenotetext{a}{In $10^{20}$ cm$^{-2}$.}
\tablenotetext{b}{Emission measure in $10^{65}$ cm$^{-3}$.}
\label{t:fits}
\end{deluxetable}

On the other hand, the soft excess apparent in the single power-law model can
also be accounted for by an extended (thus absorbed by $N_{\rm H}^{\rm G}$
only) emission of a thermal, optically-thin,  plasma (e.g., Mewe, Gronenschild
\& van den Oord 1985) at $kT_{\rm s}\sim 1$ keV. Using this model for the soft
excess yields $\chi_\nu^2= 150.5/140$, i.e., an improvement of the fit at the
98\% confidence level with respect to the model without a soft excess. The
data and the model are shown in Figure \ref{f:spectrum}. The absorbed 0.2--2
keV luminosity in the extended emission is $4.4\times 10^{42}$ erg s$^{- 1}$,
while its total unabsorbed luminosity is $9.5\times 10^{42}$ erg s$^{- 1}$,
which is only 1.8\% of the bolometric (0.1--$10^3$ keV) luminosity of
$L=5.2\times 10^{44}$ erg s$^{-1}$. A weak (and probably constant) soft X-ray
excess in 3C 120 has also been found in \exosat\/ observations by Walter \&
Courvoisier (1992), and in {\it Einstein\/} data by Turner et al.\ 
(1991).

With the model above, we find that $\Omega=0$ yields $\cnu=179.4/141$, which
corresponds to the probability of the absence of reflection of $7\times
10^{-7}$. The best-fit flux of the narrow line in our best-fit model
with reflection is null.
Also, the best-fit parameters of the broad line remain virtually the same if
we allow its flux
to be independent of reflection. Thus we hereafter
consider only the line associated with Compton reflection. The parameters of
that model are given in Table \ref{t:fits}. Note that $\wfe$ and $\Omega/2\pi$
given there are not independent but instead
related via eq.\ (\ref{eq:line}), which also yield the line
flux of $\ife= (\Omega/2\pi) 8.8\times 10^{-5}$ cm$^{-2}$ s$^{-1}$ in this
specific model. (The corresponding $\Gamma$-$\Omega$ error contour is shown in
\S \ref{s:comparison} below.)

\begin{figure}
\epsscale{1.0}
\plotone{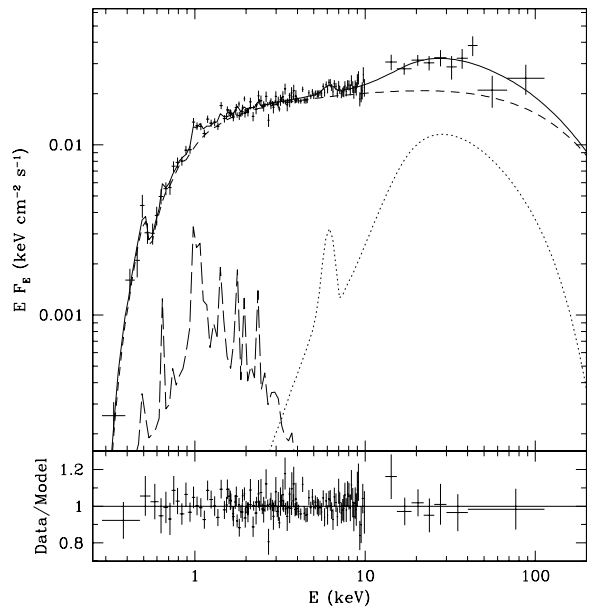}
\caption{The \sax\/ spectrum of 3C 120 (crosses; normalized to the MECS data),
modeled (solid curve) by an e-folded power law (short dashes), reflection/Fe
K$\alpha$ emission (dots) and an extended optically-thin plasma (long dashes),
see Table \ref{t:fits}. The lower panel gives the data-to-model ratio. }
\label{f:spectrum}
\end{figure}

Using this model except for allowing reflector to be ionized, we find the
reflecting surface is close to neutral, with $\xi=0^{+30}$ erg s$^{-1}$ cm,
which implies that the dominant ion is Fe {\sc xiii} (well below the region of
resonant absorption). Also, allowing the Fe abundance to vary does not improve
the fit and it has weak effect on its parameters (cf.\ E00). This confirms the
validity of of equation (\ref{eq:line}). On the other hand, theoretical
uncertainties on the Fe K$\alpha$ line flux from disk reflection (\.Zycki \&
Czerny 1994) allow it to be within $\sim 0.7$--1 of the values of George \&
Fabian (1991). If we use the lowest prediction on the disk line flux, the
allowed narrow line is still rather weak, with $\wfe= 5_{-5}^{+45}$ eV. Thus,
given the limited statitics of our data, we cannot rule out some contribution
from a remote torus, either Thomson-thin (yielding only a narrow line as in the
case above) or Thomson-thick (yielding fractions of both the observed line and
reflection). 

The Gaussian Fe K$\alpha$ line shows evidence for a broadening and redshift
(see Table \ref{t:fits}). Therefore, we also consider a model with the line
emission and Compton reflection [tied via eq.\ (\ref{eq:line})] both
originating in a relativistic disk (\S \ref{s:models}). We have found that this
model provides as good a fit, with $\cnu=151.3/142$, as the one with a Gaussian
line and static reflection. Its parameters are $\Gamma=1.86_{-0.05}^{+0.05}$,
$\Omega/2\pi=0.65_{- 0.14}^{+0.31}$, $\ec=120_{-20}^{+110}$ keV, $r_{\rm
in}=6^{+17}$, and $\wfe=100^{+40}_{-20}$ eV. With this physical model we
re-examine the significance of the presence of the soft excess. Without it, the
above model yields $\cnu=162.8/144$, which corresponds to the significance
increased 99.5\%. This shows that the physical requirement that both the line
and the reflection continuum are broadened in the same way leads to an increase
of the significance of the presence of a soft excess.

Then, we considered models with the continuum given by thermal Comptonization
(including the soft plasma emission as in Table \ref{t:fits}). We use a model
of Poutanen \& Svensson (1996), initially for spherical geometry with a
homogeneous  distribution of sources of seed photons, which we assumed to be a
blackbody at  a temperature of $kT_{\rm bb}=5$ eV. We obtain a temperature of
the scattering medium of  $kT=36^{+100}_{-6}$ keV, a Thomson optical depth of
$\tau=2.5^{+0.9}_{-1.9}$,  $\Omega/2\pi=0.56^{+0.24}_{-0.19}$, and $r_{\rm in}=
12_{-6}^{+100}$ at $\cnu= 151.0/142$. The fitted parameters weakly depend on
the assumed geometry; e.g., a geometry with hot plasma regions in the form of
cylinders on the surface of a disk yields $kT=50^{+80}_{-8}$ keV,
$\tau=2.2^{+0.3}_{-1.4}$,  $\Omega/2\pi=0.64^{+0.24}_{-0.08}$, and
$r_{\rm in}= 10_{-4}^{+4}$ at $\cnu= 151.3/142$.

The latter geometry yields an anisotropy break appearing around the peak of the
2nd scattering order (Poutanen \& Svensson 1996), which could, possibly,
account for the broken power-law spectral solution with the break at $\sim 1$
keV and no additional soft component described above. However, we have found
that a coronal geometry without an additional soft component yields a
relatively poor fit, $\cnu=159.1/144$ with $kT\simeq 120$ keV,  $\tau\simeq
0.9$. Comparing to the thermal-Compton above, the significance of the presence
of a soft component is still 97\%. If the temperature of seed photons is
allowed to be free, $kT_{\rm bb}\la 20$ eV, which constraint follows from the
anisotropy break entering the fitted energy range.

We also consider a model with a broken power-law primary continuum with a break
at high energies. However, this model requires the break to be above our data,
at $\ga 110$ keV, with the index below the break of $\Gamma = 1.91$ and an
unconstrained index above the break. This model also provides a relatively poor
fit, with $\cnu= 154.6/139$. Thus, there is no evidence for a sharp spectral
break at high energies in the data.

\section{VARIABILITY}
\label{s:var}

The spectrum analyzed in \S \ref{s:results} has been integrated over the
observation spanning more than 2 days (see Table \ref{t:log}), during which the
source varied. The LECS and MECS light curves are shown in Figure \ref{f:var}a.
In both instruments, the $\chi^2$ probability that the source were constant was
$<10^{-3}$ independently of the temporal bin used. This is in agreement with
the observed earlier on various time scales by \exosat, \rosat\/ and \asca\/
(Maraschi et al.\ 1991, hereafter M91; Grandi et al.\ 1997).

\begin{figure*}
\epsscale{2.2}
\plottwo{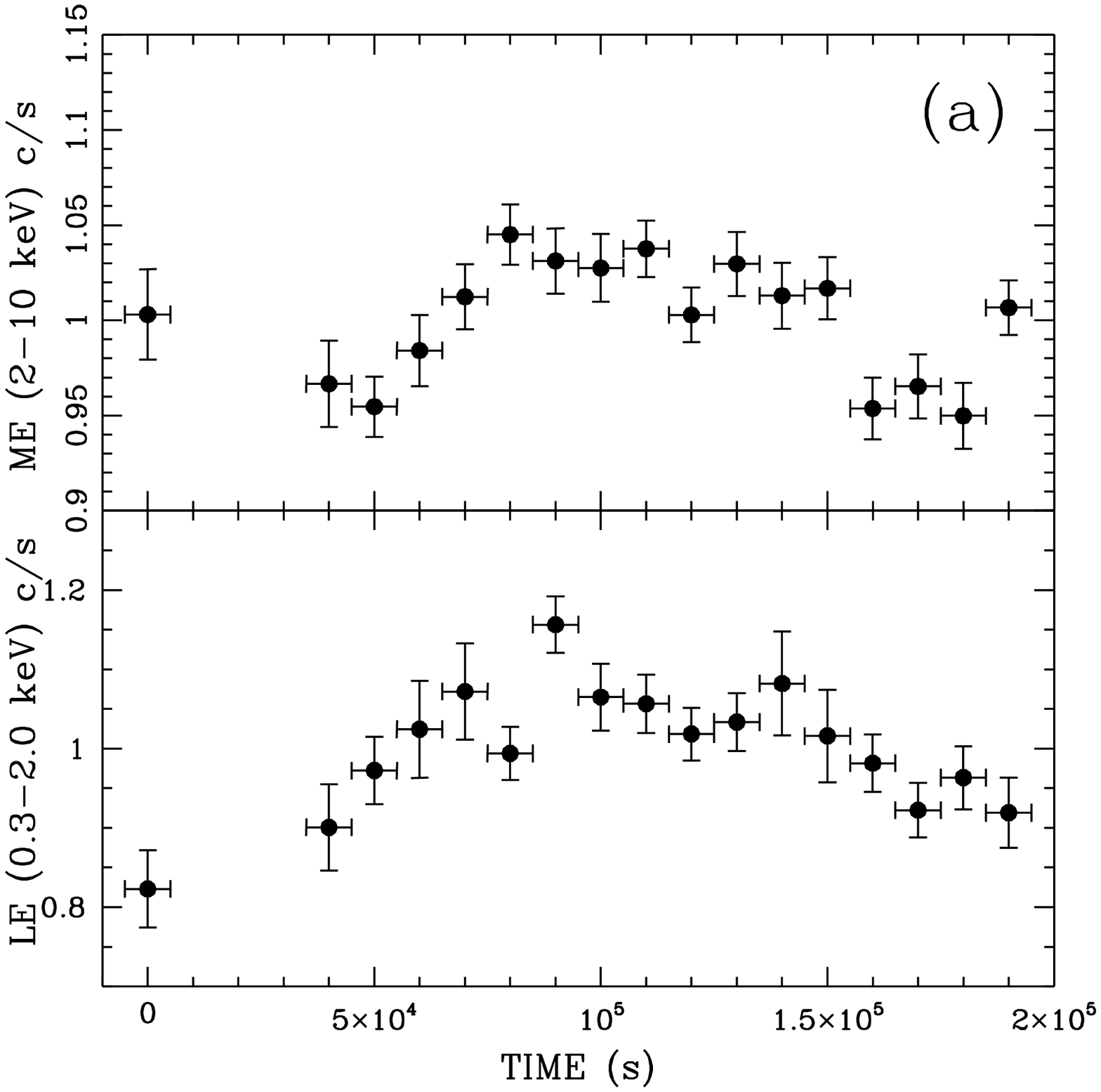}{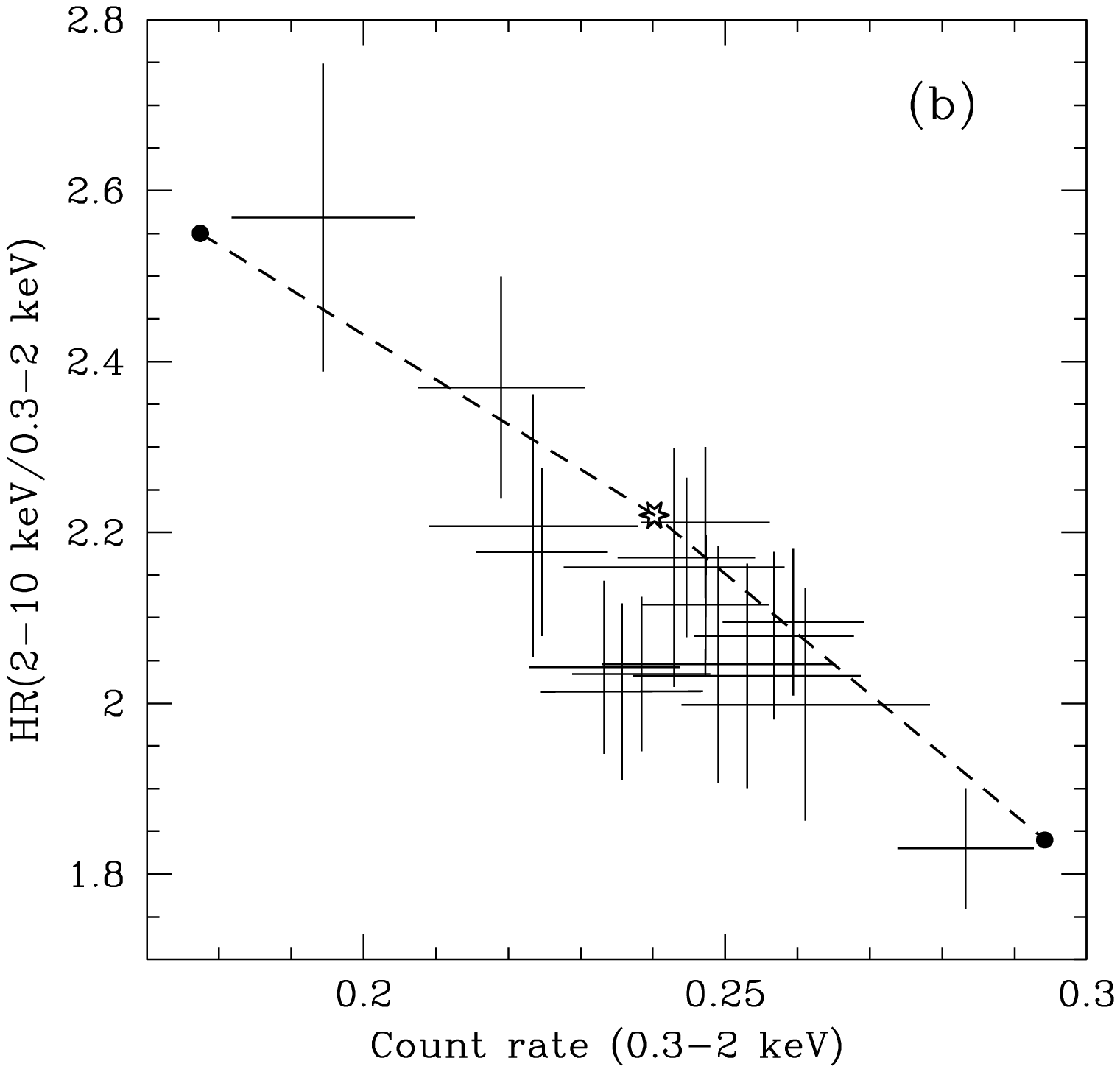}
\caption{{\it (a)\/} The LECS and MECS light curves of 3C120 normalized to the
average count rates. The bin size is $10^4$ s. {\it (b)\/} The MECS-to-LECS
count ratio (hardness ratio) as a function of the LECS count rate. The spectrum
becomes significantly softer with the increasing flux. The asterisk and the
solid circles (connected by dashed lines) correspond to the time-averaged
spectrum and the theoretical thermal-Compton spectra obtained by changing the
irradiating seed photon flux by a factor of 2, respectively.}
\label{f:var}
\end{figure*}

We see in Figure \ref{f:var}a that the soft spectrum is more variable than the
hard one. Namely, the amplitude of the variations is $\sim 15\%$ in the 0.3--2
keV energy range and only $\sim 5\%$ in the 2--10 keV band. This variability
pattern is driven by the spectrum becoming softer with increasing soft flux, as
shown in Figure \ref{f:var}b. The linear correlation coefficient between the
hardness ratio and the soft count rate is $r=0.92$, and the probability that
the 2 quantities are correlated is $>99\%$. A similar anticorrelation of the
X-ray flux and spectral index was found on longer time scales in {\it
Einstein\/} data by Halpern (1985) and in {\it EXOSAT\/} data by M91 and Grandi
et al.\ (1992).

\begin{figure}
\epsscale{1.0}
\plotone{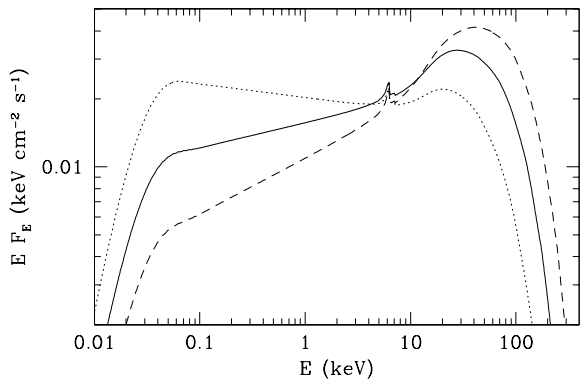}
\caption{Spectral variability of 3C 120 inferred from the variability data
shown in Fig.\ \ref{f:var}b using the thermal-Compton model. The solid
curve corresponds to the unabsorbed average spectrum (including Compton
reflection and the Fe K fluorescence but without the soft-excess
component and without the UV photons not incident on the hot plasma), and the
dashed and dotted curves correspond to changing the flux in the irradiating
soft photons by factors 1/2 and 2, respectively (which corresponds to the lower
and upper solid circle, respectively, in Fig.\ \ref{f:var}b).
}
\label{f:spec_var}
\end{figure}

In order to find the range of the spectral index, $\Gamma$, corresponding to
the anticorrelation shown in Figure \ref{f:var}b, we have used the model of
Table \ref{t:fits} in which we have varied (only) $\Gamma$ and then adjusted
the normalization to match the 0.3--2 keV LECS rate of the extreme best-fit
points in Figure \ref{f:var}b. We have found that the range of $\Gamma$
matching the observed spectral variability is quite large, 1.70--2.03
(significantly larger than the 90\%-uncertainty range of the average $\Gamma$
in Table \ref{t:fits}).

We have then investigated which physical process can be responsible for the
observed behavior. A natural candidate here is thermal Comptonization in a hot
plasma cloud irradiated by a variable flux of soft seed photons, as suggested
by M91. If the power supplied to the cloud and its Thomson
optical depth, $\tau$, are constant, its electron temperature, $kT$, will
adjust itself to the variable seed flux as to satisfy energy balance. Then, the
higher the seed flux, the lower $kT$, and (since a decrease of $kT$ at a
constant $\tau$ corresponds to an increase of the X-ray spectral index,
$\Gamma$) the softer the X-ray spectrum. Since the Comptonized spectrum joins
at low energies to the peak of the spectrum of the seed photons, this will also
correspond to an increase of the soft X-ray flux.

To test whether this process can indeed reproduce the observed spectral
variability, we have used a Comptonization code by Coppi (1992, 1999), with its
present {\sc xspec} version described in Gierli\'nski et al.\ (1999). This
model is relatively similar to the one used in \S \ref{s:results} except for
$kT$ being calculated self-consistently (rather than assumed) from  the fluxes
in  Comptonized photons, $F_{\rm C}$, and in the seed photons irradiating the
cloud, $F_{\rm seed}$. Thus, its main parameters are $\tau$, $F_{\rm C}$, and
$F_{\rm seed}$, with $F_{\rm seed}$ giving the normalization of the spectrum.
We assume that the plasma is purely thermal, and the seed photons are blackbody
at $kT_{\rm bb}=10$ eV. We then fit the time-averaged spectrum, and obtain the
amplification factor of $F_{\rm C}/F_{\rm seed} =7.4^{+2.0}_{-1.6}$ and
$\tau=2.4^{+0.3}_{-1.4}$. Reflection is tied to the Fe K line and the
relativistic smearing is taken into account as in \S \ref{s:results}, yielding
$\Omega/2\pi=0.54^{+0.15}_{-0.15}$ and $r_{\rm in}= 12_{-6}^{+250}$. We also
take into account the soft-excess emission of an optically thin plasma (\S
\ref{s:results}), and obtain $\cnu= 151.4/142$.

We then vary {\it only\/} $F_{\rm seed}$, increasing and decreasing it by a
factor of 2. Unlike the power-law model above, we do not adjust the
model normalization. The resulting predictions on the soft count rate and the
spectral hardness are shown in Figure \ref{f:var}b. Given the simplicity of our
assumption, the agreement with the observed spectral variability is extremely
good, yielding strong support to the picture of the X-ray variability driven by
a variable flux of irradiating UV photons.

The unabsorbed spectra corresponding to the inferred variability pattern are
shown in Figure \ref{f:spec_var}. Note that the shown UV and EUV part
corresponds only the blackbody photons irradiating the hot plasma, with photons
emitted directly to the observer {\it dominating\/} actual UV spectra of 3C 120
(see fig.\ 8 in M91). We see that the average and the extreme spectra pivot in
the $\sim 4$--12 keV range. This is consistent with a lack of correlation of
$\Gamma$ and the 2--10 keV flux found by Walter \& Courvoisier (1992).

A result of Walter \& Courvoisier (1992) that $\Gamma$ is approximately
proportional to the logarithm of the ratio of the UV flux to the X-ray flux is
then consistent with the variability pattern in Figure \ref{f:spec_var},
provided the UV flux incident on the hot plasma is proportional to that emitted
towards the observer. This scenario is also supported by the finding of M91
that their highest UV state corresponds to the softest X-ray spectrum.

Note that Walter \& Courvoisier (1992) have suggested that the spectral
variability in X-rays (and the UV/X-ray correlation) can be explained by
thermal Comptonization in a plasma with variable $\tau$. For our obtained
plasma parameters, we do not confirm this suggestion. Namely, varying $\tau$ by
a factor of 2 changes the soft count rate, but without any noticeable changes
in the hardness ratio, in contrast to our successful model with variable
$F_{\rm seed}$.

As we see in Figure \ref{f:spec_var}, a prediction of the thermal-Compton model
is that the lowest soft X-ray flux is associated with the highest flux in the
$\sim 20$--50 keV band as well as with the highest $kT$. Unfortunately, the
limited statistics of our PDS data do not allow us to test this prediction.
This can be tested in the future by the IBIS detector aboard {\it INTEGRAL\/}
(Ubertini et al.\ 1999). Interestingly, the spectral variability of the
black-hole binary Cyg X-1 in the hard state does show a pattern similar to that
in Figure \ref{f:spec_var}. Namely, a relatively hard spectrum with
$\Gamma\simeq 1.6$ has a high-energy cutoff corresponding to $kT\simeq 100$ keV
(Gierli\'nski et al.\ 1997) whereas a relatively soft spectrum with 
$\Gamma\simeq 2$ has $kT\simeq 60$ keV (Frontera et al.\ 2001) and the 2
spectra intersect at $\sim 5$ keV (see Zdziarski, Wen \& Paciesas 2001).

In principle, the variability shown in Figure \ref{f:var} could also be a
result of a variable $\nh$. However, setting $\nh=0$ in our models of \S
\ref{s:results} does not reproduce the highest count rate shown in Figure
\ref{f:var}. For that, we have found that the intrinsic power law index would
have to be $\Gamma\simeq 1.95$. As already found in \S \ref{s:results}, such a
$\Gamma$ gives a relatively bad fit to the average spectrum as well as it has
not been observed by \asca\/ and \xte\/ (see \S \ref{s:comparison} below).
Also, a model with variable $\nh$ has been considered by M91, who ruled it out
based on the extensive \exosat\/ monitoring. Finally, that model would not
explain the UV/X-ray correlation discussed above.

\section{COMPARISON WITH ASCA, XTE AND OSSE}
\label{s:comparison}

Figure \ref{f:comparison} shows a comparison of the spectrum from \sax\/ (black
symbols; the LECS and PDS spectra are renormalized to that of MECS) with those
from \asca\/ (red symbols; the data from the 4 \asca\/ detectors are normalized
to SIS0 and coadded, W98), \xte\/ (green symbols; the data are from E00) and
{\it CGRO}/OSSE (blue, W98).

\begin{figure*}
\epsscale{1.25}
\plotone{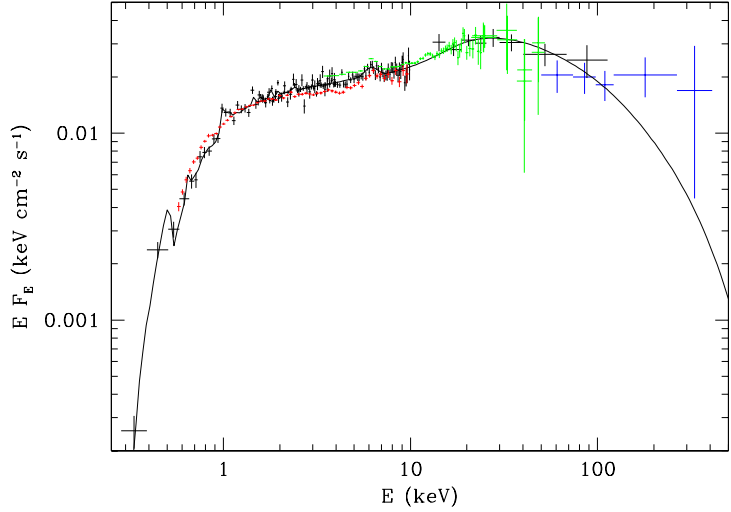}
\caption{Comparison of X-ray spectra of 3C 120 observed by \sax\/ (black
crosses and the curve), \xte\/ (green crosses), \asca\/ (red crosses) and OSSE
(blue crosses). In the 4--10 keV range, the \xte, \sax\/ and \asca\/ spectra
are from top to bottom. }
\label{f:comparison}
\end{figure*}

\begin{figure*}
\epsscale{1.2}
\plotone{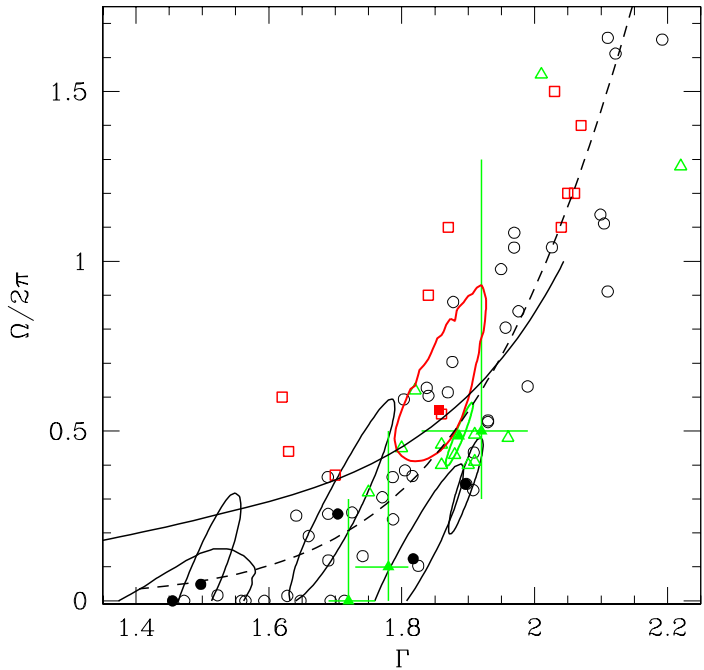}
\caption{The location of 3C 120 (heavy contours) on the $\Omega$-$\Gamma$
diagram. Circles (black), triangles (green), and squares (red) correspond to
results from \ginga, \xte, and \sax, respectively. Open and filled symbols
correspond to radio-quiet Seyferts and BLRGs, respectively. For clarity,
error contours or bars are shown only for BLRGs. The data are from Zdziarski
(1999, 2000), Matt (2000) and references therein, and E00. The solid and
dashed curves give predictions of simple models with overlapping hot and cold
disks (Z99) and a disk corona with bulk motion (Beloborodov 1999a).
}
\label{f:reflection}
\end{figure*}

As seen in Figure \ref{f:comparison}, the \asca\/ spectrum shows both the
continuum and the line profile remarkably similar to those of the \sax\/
spectrum. However, the \asca\/ data show a spectral break at $\sim 4$ keV, with
$\Gamma\simeq 2.0$ and 1.75 below and above the break, respectively (using the
same data as W98). Then the model with a broken power-law and Compton
reflection tied to the Fe K$\alpha$ line yields both the reflection and line
parameters very similar to those in the \sax\/ data, namely $\Omega/2\pi=
0.50_{-0.19}^{+0.55}$, $\efe\simeq 6.4$ keV, $\sfe\simeq 0.24$ keV, $\wfe\simeq
75_{-29}^{+83}$ eV. When the presence of the break is not taken into account
and the \asca\/ data are fitted with a single power-law continuum and a line
with free parameters, an unphysical line with $\sfe\sim \wfe\sim 1$ keV
appears, as discussed by W98.

Since the break at 4 keV is not seen in the \sax\/ data, we have considered
possible causes of its presence in the \asca\/ data. We note first that a
similar break appears in the \xte\/ data (see below), with the spectrum below
3.5 keV being significantly softer than that above it. However, this may result
from calibration problems, as it is shown to be the case, e.g., for a
simultaneous \asca/\xte\/ observation of Cyg X-1 (Gierli\'nski et al.\ 1999),
and, accordingly, E00 have not taken into account the energy range below $\sim
4$ keV in their analysis. We have then looked at possible discrepancies between
the \asca\/ SIS and GIS data. We have found that the SIS data at $\ga 5$ keV
and $\la 1$ keV are somewhat harder and softer, respectively, than the GIS data
(in both versions of the \asca\/ data of W98 and of Sambruna et al.\ 1999).
However, when we ignore the SIS data above 5 keV and either SIS or GIS data
below 1 keV, we obtain virtually identical parameters of the broken
power-law fit, with only a slight change of the hard power law of
$\Delta\Gamma < 0.05$. Thus, we conclude that the (small) discrepancy between
the SIS and GIS data is {\it not\/} the cause of the 4-keV break, or,
equivalently, of the spuriously strong and broad line appearing in fits with a
single power law continuum. We also note that in the \asca\/ sample analyzed by
W98, a break at $\sim 3$--4 keV appears also in the spectrum of 3C 382 and in
one of 3 spectra of 3C 390.3 (all observed in 1993--94) whereas is not seen in
the spectra of 3C 445, 3C 111 and the remaining two of 3C 390.3 (from
1995--96). A further analysis of this issue appears desirable.

We have also found a good agreement of the spectral shapes between the \sax\/
and \xte\/ spectra. We have refitted the \xte\/ data of E00 in the 3.5--25 keV
(PCA) and 20--50 keV (HEXTE) ranges with our baseline model (including a
thermal soft-excess component and $\nh$ with parameters fixed at those obtained
from the \sax\/ data). We find that the \xte\/ data are compatible with the
presence of an additional line component in addition to that coming from the
reflector (see \S\ \ref{s:results}). We obtain then continuum parameters
similar to those of E00, namely $\Gamma=1.89^{+0.02}_{-0.02}$,
$\Omega/2\pi=0.49^{+0.10}_{-0.10}$, yielding $\cnu=41.9/51$. The equivalent
widths of the reflection-related line and the additional line are $70\pm 15$ eV
and $15\pm 15$ eV, respectively. The data prefer the reflection-related line to
be narrow, with $\sfe= 0.09_{-0.09}^{+0.32}$ keV. In spite of these
differences, the line parameters in the \xte\/ and \sax\/ data are still
consistent with each other within 90\% confidence. The $\Gamma$-$\Omega$ error
contours for the \xte\/ data and those from \sax\/ (using the model of Table
\ref{t:fits}) are shown in Figure \ref{f:reflection}.

We also note hard tails in the PCA and HEXTE data above the energy ranges used
by us. The presence of these tails, most likely due to residual inaccuracies in
background subtraction, appear responsible for the very high e-folding energy
($\sim 10^4$ keV) and for the relatively large $\cnu=64.4/59$ obtained by
E00. When the data containing the tails are not used, $\ec$ is unconstrained by
the \xte\/ data, and we thus kept it fixed at 150 keV (obtained from the \sax\/
data).

Thus, there is an overall agreement about the parameters of both the line
and the hard X-ray continuum between the \sax, \asca\/ and \xte\/ data:
$\sfe\sim 0.2$--0.3 keV, $\wfe\sim 80$--100 eV, $\Gamma \sim 1.8$--1.9,
$\Omega/2\pi \sim 0.5$--0.6.

A somewhat unclear issue is the form of the break or cutoff at soft \g-rays.
The \xte\/ data do not allow us to determine it, as discussed above. The \sax\/
data require a high-energy break or a cutoff. In the e-folded power-law model,
shown in Figure \ref{f:comparison}, $\ec\sim 150$ keV. Also, an OSSE
observation simultaneous with that by \asca\/ required the presence of a cutoff
with $\ec\sim 100$ keV at a significance of 99.95\% (W98). However, we see that
the average OSSE data shown in Figure \ref{f:comparison} above $\sim 150$ keV
clearly lie above the predictions of this model for energies $\ga 150$ keV.
This remains the case when the OSSE data (at a free normalization) are fitted
together with the \sax\/ data, in which case the best-fit values are $\ec=250$
keV ($\cnu=185.9/178$) and $kT=100$ keV in the e-folded power-law model and the
spherical thermal-Compton model, respectively. Since the OSSE data show
the source variability at energies $\geq 50$ keV (W98), it is possible that the
cutoff energy is variable, and the spectrum at the time of the \sax\/
observation did have an exponential cutoff. On the other hand, the spectrum of
3C 120 might have a substantial high-energy tail  due to a non-thermal
continuum. This hypothesis is supported by OSSE observations of other radio
galaxies, as discussed in W98.

In particular, the OSSE data might contain a contribution from non-thermal,
Doppler-enhanced, radiation of a jet. In order to test this hypothesis, we
again fit the combined \sax/OSSE data (as above) with the model of Table
\ref{t:fits} but with an addition of a power law representing the jet emission.
This yields yields $\cnu=183.0/176$ (which corresponds to the significance of
adding the power-law component of  only 75\%) and the best-fit $\Gamma_{\rm
jet}=1.7$ (which is almost unconstrained due to the low significance of adding
that component; other best-fit parameters are $\Gamma=1.89$,
$\Omega/2\pi=0.64$, $\ec=120$ keV). The 1-keV normalization of the jet
component is $(2.3\pm 2.3) \times 10^{-3}$ cm$^{-2}$ s$^{-1}$ keV$^{-1}$ when
$\Gamma_{\rm jet}$ is fixed at 1.7. The 2 components intersect at $\sim 100$
keV. However, the integrated photon flux above 100 MeV of this model is
$(1.0\pm 1.0)\times10^{- 6}$  cm$^{-2}$ s$^{-1}$, which best-fit value is $\sim
10$ times higher than the {\it CGRO}/EGRET detection threshold. Even though
pointed several times, EGRET has not detected 3C 120 (Lin et al.\ 1993). Thus,
if such a power-law jet component exists, it has to be either very weak or have
a strong break/cutoff at energies below those observed by EGRET. On the other
hand, there is no statistically significant evidence for its existence in the
data.

\section{DISCUSSION}
\label{s:dis}

\subsection{Relationship to BLRGs and Seyferts}
\label{s:relationship}

W98 have pointed out that BLRGs appear, on average, to have less Compton
reflection, weaker F K$\alpha$ lines and harder X-ray spectra than radio-quiet
Seyfert 1s. Those conclusions have been confirmed by the studies of the \sax\/
and \xte\/ data (G00, E00).  In particular, the mean of the best-fit values of
$\Gamma$ and the standard error of measuring that mean in models with Compton
reflection in the above BLRG samples from \ginga\/ (W98), \sax\/ (G00), and
\xte\/ (E00) are $\langle \Gamma \rangle= 1.67\pm 0.09$, $1.74\pm 0.03$ and
$1.83\pm 0.08$, respectively. For the sum of the 3 samples, we obtain $\langle
\Gamma \rangle=1.74\pm 0.04$. All these values are less than the corresponding
mean for Seyfert 1s, which for the sample from \ginga\/ (Nandra \& Pounds 1994)
and \sax\/  (Matt 2000) is $\langle \Gamma \rangle= 1.95\pm 0.05$ and $1.91\pm
0.05$, respectively. (In both cases, NGC 4151, a Seyfert 1 with very strong
X-ray absorption, has not been included for computing the mean.)

These numbers clearly show that $\langle \Gamma\rangle$ in BLRGs is
significantly lower (by $\sim 5\sigma$) than that in Seyfert 1s. The
probability that the mean of the above joint sample of BLRGs equals 1.95 (i.e.,
the mean of Seyfert 1s of Nandra \& Pounds 1994) is $2.3\times 10^{-5}$
(obtained from the Student $t$ distribution). Also, we note that {\it all\/} 15
best-fit values of $\Gamma$ in the above 3 samples of BLRGs are less than
$\langle \Gamma \rangle$ of Nandra \& Pounds (1994). If $\Gamma$s of BLRGs were
sampled from the same distribution as that of Seyfert 1s, the probability of
such an event would be only $2^{-15}\simeq 3\times 10^{-5}$. Furthermore, we
can test the hypothesis that the joint sample of BLRGs has the mean of 1.95
{\it and\/} the standard deviation of 0.15 (as in the sample of Nandra \&
Pounds 1994), for which we obtain the probability of $4.7\times 10^{-8}$ (based
on the normal distribution).

We note that E00 stated that their distribution of $\Gamma$ did not differ from
that of Seyfert 1s from \ginga. However, the probability that their values of
$\Gamma$ (1.78, 1.90, 1.72, 1.92) belong to the distribution with
$\langle\Gamma \rangle=1.95$ and $\sigma_\Gamma= 0.15$ is only 5\%.
Furthermore, we should note the difference of $\Delta\Gamma \simeq 0.1$ between
\xte\/ and other X-ray instruments in fits to the Crab as well as other sources
(e.g., Gierli\'nski et al.\ 1999; Done, Madejski \& \.Zycki 2000), which, if
taken into account, would make the difference in $\langle \Gamma \rangle$
between BLRGs and Seyfert 1s even more pronounced, reducing the above
probability to $1.7\times 10^{-3}$.

On the other hand, the distribution of $\Gamma$ in BLRGs is {\it not
disjoint\/} from that in Seyfert 1s. This can be noted by considering the
intrinsic dispersion, $\sigma_\Gamma$, of the values of $\Gamma$ in the samples
(given by their standard deviations). For the BLRGs samples from \ginga, \sax\/
and \xte, $\sigma_\Gamma=0.18$, 0.08, and 0.10, respectively, while
$\sigma_\Gamma$ of their sum equals 0.14. The dispersion in the Seyfert-1
samples from \ginga\/ and \sax\/ (Nandra \& Pounds 1994; Matt 2000) is
$\sigma_\Gamma=0.15$ and 0.20, respectively. Given the values of $\langle
\Gamma \rangle$ listed above, the distributions for BLRGs and Seyfert 1s
significantly overlap (which, we stress again, does not imply that the 2
distributions are indistinguishable).

Furthermore, BLRGs obey an overall correlation between these  $\Gamma$ and
$\Omega$ seen in Seyferts (Zdziarski, Lubi\'nski \& Smith 1999, hereafter Z99).
This is illustrated in Figure \ref{f:reflection} (where we can also note the
$\Delta\Gamma\sim 0.1$ offset between the \xte\/ results and those of \ginga\/
and \sax). We see that BLRGs (including 3C 120) simply occupy a lower left part
of the broader $\Gamma$-$\Omega$ parameter space occupied by radio-quiet
Seyferts. Also, the average strength of the broad component of the Fe K$\alpha$
line in radio-quiet Seyferts with $\Gamma$ close to that of 3C 120 observed by
\asca\/ (Lubi\'nski \& Zdziarski 2001) corresponds to the strength of Compton
reflection consistent within error bars with that observed in 3C 120.

This implies a general similarity between the physical conditions in the X-ray 
sources of Seyferts and BLRGs (including 3C 120), but with some unknown 
parameter determining ($\Gamma$, $\Omega$) in BLRGs offset with respect to that 
in radio-quiet Seyferts. We discuss this issue in the \S \ref{s:geometry} 
below.

The overall similarity between the X-ray properties of BLRGs and radio-quiet
Seyferts is further confirmed by the variability pattern of 3C 120 (Fig.\
\ref{f:var}b), where the X-ray spectrum becomes softer with the increasing
X-ray flux. This pattern is the same as that often seen in radio-quiet Seyfert
1s, e.g., in NGC 5548 (Magdziarz et al.\ 1998) and NGC 4151 (Yaqoob \& Warwick
1991). We also note that black-hole binaries in the hard state show a similar
behavior (e.g., Zdziarski, Wen \& Paciesas 2001).

\subsection{Radiative and pair processes}
\label{s:rad}

Clearly, the dominant radiative process responsible for X-ray continua of many
classes of compact objects is Compton upscattering of soft photons by energetic
electrons. Less clear issues are the form of the electron distribution and the
source of the seed soft photons. In the case of radio-quiet Seyfert 1s, the
form of the soft \g-ray spectra from OSSE rules out distributions dominated by
non-thermal electrons (Gondek et al.\ 1996). The strongest evidence for an
electron distribution close to Maxwellian with $kT\sim 50$--100 keV and
$\tau\sim 1$ in Seyferts comes from the average OSSE spectrum of NGC 4151
(Johnson et al.\ 1997) and the combined average OSSE spectra of other Seyferts
(Zdziarski, Poutanen \& Johnson 2000).

In the case of 3C 120 (as well as other BLRGs, W98), there is a break
or a cutoff in soft \g-rays, but its form is not well determined. The
thermal-Compton model gives $kT\sim 30$--130 keV, but the OSSE data above 100
keV lie somewhat above the predictions of this model, indicating a possible
high-energy, non-thermal, tail.

On the other hand, the fact that the X-ray spectra of BLRGs obey the
$\Omega$-$\Gamma$ correlation supports their predominantly thermal origin. As
discussed in Z99, the most likely origin of the correlation is a feedback
between cold and hot media. The cold medium Compton reflects hard emission and
provides blackbody seed photons for thermal-Compton cooling of the hot, X-ray
emitting, plasma. (If the electrons in the hot plasma were predominantly
non-thermal, no relation of the X-ray slope to the rate of cooling would be
required.) Furthermore, the observed softening of the X-ray spectrum with
increasing flux (Figs.\ \ref{f:var}-\ref{f:spec_var}) strongly supports the
thermal-Compton model with a variable flux of irradiating soft photons, as
shown in \S \ref{s:var}.

As stated above, the presence of the $\Omega$-$\Gamma$ correlation implies that
the seed photons Comptonized in the hot plasma are emitted by a surrounding
cold medium rather than intrinsically to the hot plasma. This is independently
supported by results of Wardzi\'nski \& Zdziarski (2000, see their fig.\ 8),
who found that the expected strength of magnetic field in accreting sources in
luminous AGNs (such as 3C 120) is far too weak to provide enough seed photons
via thermal synchrotron emission for Comptonization to account for the observed
X-ray spectra.

The best-fit thermal-Compton models (\S \ref{s:results}) have $kT$ ($\simeq
40$--50 keV) too low for the plasma clouds with $\tau\sim 2$ and sizes $\ga
R_{\rm g}$ to be made predominantly of \ee\ pairs, as follows from
consideration of equilibrium between pair production and annihilation (Stern et
al.\ 1995; Svensson 1996). Pair-dominated plasmas are only marginally possible
for the highest $kT$/lowest $\tau$ compatible with the data at 90\% confidence,
$kT\simeq 130$ keV, $\tau<1$, provided the active regions are small enough
(Svensson 1996). On the other hand, a modest non-thermal tail in the electron
distribution could greatly enhance the pair-production rate, easing the above
constraint. Thus, our data do not significantly constrain the pair content in
the X-ray emitting plasma in 3C 120. We note that even if the plasma is
pair-dominated, the presence of a distinct \ee\ annihilation feature around 511
keV is unlikely (Macio{\l}ek-Nied\'zwiecki, Zdziarski \& Coppi 1995). To
conclusively resolve the issue of the presence of pairs, high signal-to-noise 
data up to $\sim 1$ MeV (e.g., from the IBIS/{\it INTEGRAL\/} detector) are
required.

\subsection{The geometry of the X-ray source}
\label{s:geometry}

Two main disk geometries have been invoked for accreting X-ray sources: a
patchy corona above a cold accretion disk, and a hot accretion disk with an
overlapping cold medium (either an outer cold disk, cold blobs or both). As we
show below, both of those possibilities appear viable for 3C 120, and BLRGs in
general.

In the case of coronal geometry, the values of $kT$ and $\Gamma$ fitted to the
\sax\/ spectrum of 3C 120 (\S \ref{s:results}) strongly rule out either a
homogeneous corona or active regions located directly on the disk surface, from
considerations of energy and pair balance (Stern et al.\ 1995). On the other
hand, models with static active regions elevated above the disk to a height
comparable to their size can fit 3C 120, see Svensson (1996). In this geometry,
the fitted $\Omega<2\pi$ can be explained by a partial obscuration of the
reflecting disk surface by the active regions.

However, in order to produce harder spectra, like those of other BLRGs, the
emitting region  should be even more separated from the cold accretion disk
(e.g., Svensson 1996), but in that case the predicted $\Omega/2\pi \sim 1$ and
$\Omega$ would increase with decreasing $\Gamma$, {\it opposite\/} to
the observed $\Gamma$-$\Omega$ correlation (Fig.\ \ref{f:reflection}). This
problem can be solved by mildly relativistic motion of emitting plasma away
from the disk, in which case both hard spectra and little reflection are
obtained (Beloborodov 1999a, b). The dashed curve in Figure \ref{f:reflection}
shows a specific version of this model, as computed in Z99. We see that it
can approximately reproduce the values of $\Gamma$ and $\Omega$ of BLRGs.

This model has been claimed to be untenable for BLRGs by E00 on the grounds
that radio jets in BLRGs are highly relativistic, with $\gamma\sim 10$.
However, radio jets (in particular the relativistic jet in 3C 120, Walker
1997) are observed on scales much larger than $r\sim 10$--100 inferred
from the observed Fe K$\alpha$ line broadening. The process of jet formation
remains not well understood, but it is quite likely that the matter in the jet
originates in an accretion disk. In that case, there has to be an inner region
with a mildly relativistic wind velocity before $\gamma\sim 10$ is reached.
That inner region may be then responsible for the observed X-ray emission.

The main alternative model involves a hot accretion disk (Shapiro, Lightman \&
Eardley 1976). Advection in the flow results in a branch with ineffective
cooling as well as it limits the possible maximum luminosity and the optical
depth of an inner flow (Narayan \& Yi 1995; Abramowicz et al.\ 1995). The
strength of reflection depends on the amount of overlap of the hot disk with an
outer cold disk. The less overlap, the less effective cooling by blackbody
photons emitted by the cold disk and the harder the X-ray spectrum. Results of
a simple calculation of this effect by Z99 are shown by the solid curve in
Figure \ref{f:reflection}. We see that this model is also roughly capable of
explaining the data for 3C 120 and most other BLRGs. In the case of 3C 120, the
overlap is required by the relatively large fitted value of $\Omega$ since a
sharp radial transition between the hot and cold phases yields $\Omega/2\pi \la
0.4$ (see Z99).

Both models explain the anticorrelation between the soft X-ray flux and the
spectral hardness as being due to the increased cooling of the hot plasma by
emission of the cold disk. In the hot accretion disk model, this can naturally
be associated with the cold disk moving inward. In the outflow model, an
increased soft flux can be directly due to an increase of the accretion rate.

The fact that BLRGs have harder spectra with less reflection than those average
for Seyfert 1s indicates that some physical parameter (presumably the same that
leads to the strong radio emission and jet formation in BLRGs) leads (on
average) either to stronger coronal outflows (in the disk-corona model) or to
larger inner radii of the cold disk (in the hot/cold disk model) than in the
case of radio-quiet Seyferts.

\subsection{The soft component}
\label{s:soft}

In \S \ref{s:results}, we have compared 3 pairs of models with and without an
additional soft component. In the case of static reflection and a Gaussian
line, the fit improvement due to adding emission of an optically-thin thermal
plasma has 98\% significance. However, the model without that component has a
line with a large width incompatible with the assumed static reflection. When
both the line and reflection are relativistically smeared, this significance
increases to 99.5\%. Finally, this significance for coronal Comptonization
regions (which soft X-ray spectrum differs from a power law) is 97\%. Using the
last model, we have also demonstrated that the broken-power law fit in \S
\ref{s:results} cannot be physically justified as an approximation to the
Comptonization anisotropy break. Thus, combining our spectral analysis with
physical constraints we find the presence of a soft X-ray excess in the data to
be confirmed at a high significance.

The optically-thin emission is likely to come from a larger, kpc-scale, region.
Extended thermal components are  commonly detected in soft X-rays from 3C radio
galaxies, especially of the FR I type (Hardcastle \& Worrall 1999). A recent
analysis of \rosat\/ data has shown the presence of a hot gaseous environment
in the BLRG 3C 382 (Prieto 2000), which X-ray spectrum is rather similar to
that of 3C 120 (W98, G00). In fact, a re-analysis of the \rosat/HRI image of 3C
120 (D. Harris, private communication) appears to show an extended halo, with
the 0.2--2 keV luminosity emitted beyond the core radius of $10''$ (6.5 kpc)
comparable to that emitted by the optically-thin plasma in our spectral model
(\S \ref{s:results}). This result strongly supports the actual presence of the
soft excess in 3C 120 (also found in \exosat\/ data by Walter \& Courvoisier
1992).

Alternatively, the optically-thin emission can come from an intercloud medium
within the broad-line region. The emission measure obtained by us is, in fact,
consistent with the typical intercloud density ($\sim 10^7$ cm$^{-3}$) and the
size of the broad-line region in 3C 120 of $\sim 43$ light-days, inferred from
reverberation techniques (Wandel, Peterson \& Malkan 1999).

A somewhat worse fit was obtained with an additional steep power law, which
could represent emission from a soft X-ray jet. However, the only resolved
feature away ($25''$) from the nucleus in the \rosat/HRI image of 3C 120
(Harris et al.\ 1999) emits a negligible soft X-ray luminosity, $\sim 10^{-3}$
of the total X-ray/$\gamma$-ray luminosity, which makes this model relatively
unlikely.

Also, the soft power law fitted by us could also represent a high-energy
extension of the UV bump present in 3C120 (M91). This, however, requires rather
high temperatures of the UV emitting medium, which could, in principle, be
achieved in a transition layer between the disk and the corona (e.g., Czerny \&
Elvis 1987). This (together with the observed variability pattern, \S
\ref{s:var}) would require the soft X-ray excess to vary together with the main
X-ray component, which may be in conflict with results of Walter \& Courvoisier
(1992).

Finally, we note that the pattern of spectral variability shown in Figure
\ref{f:spec_var} shows a variable power law index. The resulting average
spectrum will be concave as the softest and hardest power law will dominate in
the lowest and highest energies, respectively. This concave spectrum can, in
principle, be responsible for the fitted soft excess.

Thus, we cannot conclusively determine the nature of the soft component due to
the limited X-ray imaging and spectroscopic capabilities of \sax. To achieve
this goal, observations with {\it Chandra\/} and {\it XMM\/} are highly
desirable.

\section{CONCLUSIONS}
\label{s:con}

Our \sax\/ data clearly show a continuum due to reflection of an X-ray power
law with $\Gamma\sim 1.8$--1.9 from a cold material. The significance of the
presence of reflection is $> 99.99\%$ and the solid angle of the reflector is
$\la 2\pi$ ($\Omega/2\pi= 0.56_{-0.20}^{+0.44}$ for the e-folded power-law
incident continuum). The reflection is accompanied by a fluorescent Fe
K$\alpha$ line with an equivalent width ($\sim 80$ eV) fully consistent with
originating entirely from the reflector. The line is moderately broad, and
combining our spectral constraints with those from \asca\/ and \xte\/ (which
data show an overall agreement with those of \sax) yields $\sfe\simeq 0.2$--0.3
keV. If this line comes from a Keplerian disk in a Schwarzschild metric, the
inner radius of the emitting region is $\ga 10GM/c^2$.

We find the presence of a high-energy break or cutoff in the \sax\/ data. We
can fit the data by either an e-folded power law or thermal Comptonization,
obtaining $E_{\rm c}= 150_{-30}^{+230}$ keV, $kT\sim 40$--130 keV,
respectively. In the latter model, $\tau\sim 0.6$--3. We also detect a
relatively weak (but significant at a $\sim 98\%$ confidence) soft excess. It
is best-fitted by emission from an extended thermal plasma at $kT_{\rm s}\simeq
1$ keV, which model is supported by an independent finding of an extended soft
X-ray halo from 3C 120.

We find strong spectral variability during the 2-day observation, corresponding
to  $\Gamma$ varying from $\sim 1.7$ to $\sim 2$ and correlated with the soft
X-ray flux. The variability is very well modeled by thermal Comptonization in a
hot plasma with a constant $\tau$ which temperature adjusts to a variable flux
of irradiating UV photons.

The 3C 120 data obey the general correlation between the power-law slope and
the strength of Compton reflection found in Seyferts and black-hole binaries.
Interpretation of this correlation involves feedback between a thermal
Comptonizing plasma and a cold medium, yielding support for our thermal-Compton
spectral model. Our data are consistent with two geometries: an outflow above
the surface of a disk, or a hot inner disk surrounded by an overlapping cold
disk. Both homogeneous and static patchy corona models can be ruled out.

We confirm (at a very high confidence level) that BLRGs have harder X-ray
spectra, less Compton reflection and weaker Fe K$\alpha$ lines than those
average in radio-quiet Seyferts. However, the distributions of those parameters
in BLRGs and radio-quiet Seyferts are not disjoint; the former occupy a part of
the parameter space occupied by the latter.

\acknowledgements

AAZ has been supported in part by a grant from the Foundation for Polish
Science and KBN grants 2P03D00614 and 2P03C00619p0(1,2). We are grateful to P.
Wo\'zniak for help in preparing the proposal for the observation analyzed here
and E. Piconcelli for partially reducing the data. We also thank M.
Gierli\'nski for help with {\sc xspec} models, M. Eracleous and R. Sambruna for
sending us their \xte\/ and \asca\/ data, D. Harris for letting us know of his
results prior to publication, K. Leighly, G. Malaguti, G. Palumbo and J.
Poutanen for valuable discussions, and the anonymous referee for insightful
comments.

\end{document}